\documentclass[a4paper,11pt]{article}
\usepackage[utf8x]{inputenc}
\usepackage[lmargin=2.5cm,rmargin=2.5cm,tmargin=2.5cm,bmargin=3.5cm]{geometry}
\usepackage[font=small,labelfont=bf,width=0.95\textwidth]{caption}
\usepackage{cite}
\usepackage{subfigure}
\usepackage{amsmath}
\usepackage{amssymb}
\usepackage{fixltx2e}             
\usepackage[T1]{fontenc}          
\usepackage{graphicx}             
\usepackage{feynmp}
\usepackage{multicol}
\usepackage{etoolbox}
\patchcmd{\thebibliography}{\section*{\refname}}
    {\begin{multicols}{2}[\section*{\refname}]}{}{}
\patchcmd{\endthebibliography}{\endlist}{\endlist\end{multicols}}{}{}
\usepackage[pdftitle={Renormalization of vacuum expectation values in spontaneously broken gauge theories},
pdfauthor={Marcus Sperling,Dominik Stockinger,Alexander Voigt},
pdfkeywords={Renormalization,Vacuum,VEV,MSSM,NMSSM,E6SSM}, bookmarks=true, linktocpage]{hyperref}

\DeclareGraphicsExtensions{.pdf}  

\newcommand{\fmfvcenter}[1]{\vcenter{\hbox{\fmfreuse{#1}}}}	

\DeclareMathOperator{\tr}{\mathrm{Tr}}			

\DeclareMathAlphabet{\mathitbf}{OML}{cmm}{b}{it}	

\newcommand{\GenGauge}[3]{T_{#1 #2}^{#3}}		
\newcommand{\CasiGauge}[2]{C_{#1 #2}^{2}(\mathrm{S})} 	

\newcommand{\GenSpiGauge}[3]{t_{#1 #2}^{#3}}			

\newcommand{\YukaGauge}[3]{Y_{#1 #2}^{#3}}			
\newcommand{\YukaGaugeConj}[3]{Y_{#1 #2}^{* #3}}		
\newcommand{\YukaInvGauge}[2]{Y_{#1 #2}^{2}(\mathrm{S})}	

\newcommand{\MassScaGauge}[2]{m_{#1 #2}^2}			
\newcommand{\TriScaGauge}[3]{h_{#1 #2 #3}}			
\newcommand{\QuadScaGauge}[4]{\lambda_{#1 #2 #3 #4}}		

\newcommand{\VecGaugeDo}[2]{V_{#2}^{#1}}		
\newcommand{\ScaGauge}[1]{\varphi_{#1}}			
\newcommand{\ScaBackGauge}[1]{\hat{\varphi}_{#1}}	
\newcommand{\BrstBackGauge}[1]{\hat{q}_{#1}}		
\newcommand{\VevGauge}[1]{\hat{v}_{#1}}			
\newcommand{\GhoGauge}[1]{c^{#1}}			
\newcommand{\AntiGhoGauge}[1]{\bar{c}^{#1}}		
\newcommand{\NaLauGauge}[1]{B^{#1}}			

\newcommand{\Spinor}[1]{\psi_{#1}}				
\newcommand{\SpinorDo}[2]{\psi_{#1 #2}}				
\newcommand{\SpinorUp}[2]{\psi_{#1}^{#2}}			
\newcommand{\AdjSpinorUp}[2]{\bar{\psi}_{#1}^{\dot{#2}}}	
\newcommand{\MassSpinor}[2]{\left(m_f\right)_{#1 #2}}		

\newcommand{\ESSM}{E\texorpdfstring{\textsubscript{6}}{6}SSM}
\newcommand{\MSbar}{\ensuremath{\overline{\text{MS}}}}
\newcommand{\DRed}{\ensuremath{\overline{\text{DR}}}}

\title{\bf Renormalization of vacuum expectation values in
  spontaneously broken gauge theories}
\author{Marcus Sperling, Dominik St\"ockinger, Alexander Voigt\\[2em]
{\sl Institut f\"ur Kern- und Teilchenphysik,
TU Dresden, Dresden, Germany}}
\date{}

\begin{document}


\newsavebox{\feynmanrules}
\sbox{\feynmanrules}{
\begin{fmffile}{Feynman/FeynmanRules} 
  \fmfset{thin}{.8pt}
  \fmfset{wiggly_len}{2mm}
  \fmfset{dash_len}{2.5mm}
  \fmfset{dot_size}{1thick}
  \fmfset{arrow_len}{2.0mm}
  \fmfset{curly_len}{2.5mm}
  \fmfset{dot_len}{1.6pt}

\begin{fmfgraph}(80,60)
  \fmfkeep{ScaSelfCT}
  \fmfleft{v1}
  \fmfright{v2}
  \fmf{dashes}{v1,c}
  \fmf{dashes}{v2,c}
  \fmfct{c}
\end{fmfgraph}

\begin{fmfgraph}(80,60)
  \fmfkeep{ScaSelfBlob}
  \fmfleft{v1}
  \fmfright{v2}
  \fmf{dashes}{v1,c}
  \fmf{dashes}{c,v2}
  \fmfv{decor.shape=circle,decor.filled=shaded,decor.size=20}{c}
\end{fmfgraph}

\begin{fmfgraph*}(80,60)
  \fmfkeep{TwoPtCTQandBRST}
  \fmfleft{v1}
  \fmfright{v2}
  \fmf{dbl_dashes,label=$\BrstBackGauge{a}$,label.side=left}{v1,c1}
  \fmf{dashes,label=$K_{\ScaGauge{b}}$,label.side=left}{c2,v2}
  \fmf{phantom,tension=5}{c1,c2}
  \fmfct{c1}  
  \fmfv{decor.shape=square,decor.filled=empty,decor.size=8}{c2}
\end{fmfgraph*}

\begin{fmfgraph*}(80,60)
  \fmfkeep{TwoPt2LCTQandBRST}
  \fmfleft{v1}
  \fmfright{v2}
  \fmf{dbl_dashes,label=$\BrstBackGauge{a}$,label.side=left}{v1,c1}
  \fmf{dashes,label=$K_{\ScaGauge{b}}$,label.side=left}{c2,v2}
  \fmf{phantom,tension=5,label=$\delta\hat{Z}^{(2)}$,label.side=right}{c1,c2}
  \fmfv{decor.shape=square,decor.filled=empty,decor.size=8}{c2}
  \fmfct{c1} 
\end{fmfgraph*}

\begin{fmfgraph*}(100,75)
  \fmfkeep{TwoPtSca1Gh1}
  \fmfleft{v1}

  \fmfright{v2}
  \fmf{dbl_dashes,label=$\BrstBackGauge{a}$,label.side=left}{v1,c1}
  \fmf{dashes,label=$K_{\ScaGauge{b}}$,label.side=left}{c2,v2}
  \fmfv{decor.shape=square,decor.filled=empty,decor.size=8}{c2}
  \fmf{ghost,left,tension=0.3}{c1,c2}
  \fmf{dashes,right,tension=0.3}{c1,c2}
  \fmfdot{c1}
\end{fmfgraph*}

\begin{fmfgraph*}(100,75)
    \fmfkeep{TwoPt2LSca1Gh1Spi2}
    \fmfleft{v1}
    \fmfright{v2}
    \fmf{dbl_dashes,label=$\BrstBackGauge{a}$,label.side=left}{v1,c1}
    \fmf{dashes,label=$K_{\ScaGauge{b}}$,label.side=left}{c2,v2}
    \fmf{ghost,left,tension=0.3}{c1,c2}
    \fmf{dashes,tension=1}{c1,c3}
    \fmf{dashes,tension=1}{c4,c2}
    \fmf{plain,left,tension=0.5}{c3,c4}
    \fmf{plain,right,tension=0.5}{c3,c4}
    \fmfv{decor.shape=square,decor.filled=empty,decor.size=8}{c2}
    \fmfdot{c1,c3,c4}
\end{fmfgraph*}

\begin{fmfgraph*}(80,60)
    \fmfkeep{TwoPtCT2LSca1Gh1Spi2}
    \fmfleft{v1}
    \fmfright{v2}
    \fmf{dbl_dashes,label=$\BrstBackGauge{a}$,label.side=left}{v1,c1}
    \fmf{dashes,label=$K_{\ScaGauge{b}}$,label.side=left}{c2,v2}
    \fmf{ghost,left,tension=0.3}{c1,c2}
    \fmf{dashes,tension=0.4}{c1,c3}
    \fmf{dashes,tension=0.4}{c4,c2}
    \fmfv{decor.shape=square,decor.filled=empty,decor.size=8}{c2}
    \fmfdot{c1}
    \fmfct{c3}
    \fmf{phantom,tension=10,label=$\delta Z^{(1)}$,label.side=right}{c3,c4}
\end{fmfgraph*}

\begin{fmfgraph*}(80,60)
  \fmfkeep{ThreePtBrstSource1Sca1Gh1}
  \fmfleft{v1}
  \fmfright{v2,v3}
  \fmf{dashes,label=$K_{\ScaGauge{a}}$,label.side=left}{v1,c}
  \fmfv{decor.shape=square,decor.filled=empty,decor.size=8}{c}
  \fmf{ghost,label=$\GhoGauge{A}$,label.side=right}{v3,c}
  \fmf{dashes,label=$\ScaGauge{b}$,label.side=left}{c,v2}
\end{fmfgraph*}

\begin{fmfgraph*}(80,60)
  \fmfkeep{ThreePtBrstBack1Sca1AntiGh1}
  \fmfleft{v1}
  \fmfright{v2,v3}
  \fmf{dbl_dashes,label=$\BrstBackGauge{a}$,label.side=left}{v1,c}
  \fmf{ghost,label=$\AntiGhoGauge{A}$,label.side=left}{c,v3}
  \fmf{dashes,label=$\ScaGauge{b}$,label.side=left}{c,v2}
  \fmfdot{c}
\end{fmfgraph*}

\begin{fmfgraph*}(80,60)
  \fmfkeep{TwoPtBrstBackTwoLoopScaSelf}
  \fmfleft{i}
  \fmfright{o}
  \fmf{dbl_dashes,label=$\BrstBackGauge{a}$,label.side=left}{i,v1}
  \fmf{dashes,label=$K_{\ScaGauge{b}}$,label.side=left}{v2,o}
  \fmfdot{v1}
  \fmfv{decor.shape=square,decor.filled=empty,decor.size=8}{v2}
  \fmf{dashes,tension=0.3}{v1,v3}
  \fmf{dashes,tension=0.3}{v3,v2}
  \fmfv{decor.shape=circle,decor.filled=shaded,decor.size=15}{v3}
  \fmf{ghost,left,tension=0.3}{v1,v2}
\end{fmfgraph*}

\begin{fmfgraph*}(80,60)
  \fmfkeep{TwoPtBrstBackTwoLoopScaSelfCT}
  \fmfleft{i}
  \fmfright{o}
  \fmf{dbl_dashes,label=$\BrstBackGauge{a}$,label.side=left}{i,v1}
  \fmf{dashes,label=$K_{\ScaGauge{b}}$,label.side=left}{v2,o}
  \fmfdot{v1}
  \fmfv{decor.shape=square,decor.filled=empty,decor.size=8}{v2}
  \fmf{dashes,tension=0.3}{v1,v3}
  \fmf{dashes,tension=0.3}{v3,v2}
  \fmfct{v3}
  \fmf{ghost,left,tension=0.3}{v1,v2}
\end{fmfgraph*}

\begin{fmfgraph*}(80,60)
  \fmfkeep{TwoPtBrstBackTwoLoopGhoSelf}
  \fmfleft{i}
  \fmfright{o}
  \fmf{dbl_dashes,label=$\BrstBackGauge{a}$,label.side=left}{i,v1}
  \fmf{dashes,label=$K_{\ScaGauge{b}}$,label.side=left}{v2,o}
  \fmfdot{v1}
  \fmfv{decor.shape=square,decor.filled=empty,decor.size=8}{v2}
  \fmf{ghost,tension=0.1}{v1,v3}
  \fmf{ghost,tension=0.1}{v3,v2}
  \fmfv{decor.shape=circle,decor.filled=shaded,decor.size=15}{v3}
  \fmf{dashes,left,tension=0.3}{v1,v2}
\end{fmfgraph*}

\begin{fmfgraph*}(80,60)
  \fmfkeep{TwoPtBrstBackTwoLoopGhoSelfCT}
  \fmfleft{i}
  \fmfright{o}
  \fmf{dbl_dashes,label=$\BrstBackGauge{a}$,label.side=left}{i,v1}
  \fmf{dashes,label=$K_{\ScaGauge{b}}$,label.side=left}{v2,o}
  \fmfdot{v1}
  \fmfv{decor.shape=square,decor.filled=empty,decor.size=8}{v2}
  \fmf{ghost,tension=0.1}{v1,v3}
  \fmf{ghost,tension=0.1}{v3,v2}
  \fmfct{v3}
  \fmf{dashes,left,tension=0.3}{v1,v2}
\end{fmfgraph*}

\begin{fmfgraph*}(80,60)
  \fmfkeep{TwoPtBrstBackTwoLoopVecExchange}
  \fmfleft{i}
  \fmfright{o}
  \fmftop{t}
  \fmfbottom{b}
  \fmf{phantom,tension=1}{t,v3}
  \fmf{phantom,tension=1}{b,v4}
  \fmf{dbl_dashes,label=$\BrstBackGauge{a}$,label.side=left}{i,v1}
  \fmf{dashes,label=$K_{\ScaGauge{b}}$,label.side=left}{v2,o}
  \fmfdot{v1}
  \fmfv{decor.shape=square,decor.filled=empty,decor.size=8}{v2}
  \fmf{ghost,tension=0.3}{v1,v3}
  \fmf{ghost,tension=0.3}{v3,v2}
  \fmf{dashes,tension=0.3}{v1,v4}
  \fmf{dashes,tension=0.3}{v4,v2}
  \fmf{photon,tension=0}{v3,v4}
  \fmfdot{v1,v3,v4}
\end{fmfgraph*}

\begin{fmfgraph*}(80,60)
  \fmfkeep{TwoPtBrstBackTwoLoopScaExchange}
  \fmfleft{i}
  \fmfright{o}
  \fmftop{t}
  \fmfbottom{b}
  \fmf{phantom,tension=1}{t,v3}
  \fmf{phantom,tension=1}{b,v4}
  \fmf{dbl_dashes,label=$\BrstBackGauge{a}$,label.side=left}{i,v1}
  \fmf{dashes,label=$K_{\ScaGauge{b}}$,label.side=left}{v2,o}
  \fmfdot{v1}
  \fmfv{decor.shape=square,decor.filled=empty,decor.size=8}{v2}
  \fmf{ghost,tension=0.3}{v1,v3}
  \fmf{ghost,tension=0.3}{v3,v2}
  \fmf{dashes,tension=0.3}{v1,v4}
  \fmf{dashes,tension=0.3}{v4,v2}
  \fmf{dashes,tension=0}{v3,v4}
  \fmfdot{v1,v3,v4}
\end{fmfgraph*}

\begin{fmfgraph*}(80,60)
  \fmfkeep{TwoPtBrstBackTwoLoopGhoExchange}
  \fmfleft{i}
  \fmfright{o}
  \fmftop{t}
  \fmfbottom{b}
  \fmf{phantom,tension=1}{t,v3}
  \fmf{phantom,tension=1}{b,v4}
  \fmf{dbl_dashes,label=$\BrstBackGauge{a}$,label.side=left}{i,v1}
  \fmf{dashes,label=$K_{\ScaGauge{b}}$,label.side=left}{v2,o}
  \fmfdot{v1}
  \fmfv{decor.shape=square,decor.filled=empty,decor.size=8}{v2}
  \fmf{ghost,tension=0.3}{v1,v3}
  \fmf{ghost,tension=0.3}{v4,v2}
  \fmf{dashes,tension=0.3}{v1,v4}
  \fmf{dashes,tension=0.3}{v3,v2}
  \fmf{ghost,tension=0}{v3,v4}
  \fmfdot{v1,v3,v4}
\end{fmfgraph*}

\begin{fmfgraph*}(80,60)
  \fmfkeep{TwoPtBrstBackTwoLoopCTqVertex}
  \fmfleft{i}
  \fmfright{o}
  \fmf{dbl_dashes,label=$\BrstBackGauge{a}$,label.side=left}{i,v1}
  \fmf{dashes,label=$K_{\ScaGauge{b}}$,label.side=left}{v2,o}
  \fmfct{v1}
  \fmfv{decor.shape=square,decor.filled=empty,decor.size=8}{v2}
  \fmf{ghost,left,tension=0.3}{v1,v2}
  \fmf{dashes,right,tension=0.3}{v1,v2}
\end{fmfgraph*}

\begin{fmfgraph*}(80,60)
  \fmfkeep{TwoPtBrstBackTwoLoopCTbrstVertex}
  \fmfleft{i}
  \fmfright{o}
  \fmf{dbl_dashes,label=$\BrstBackGauge{a}$,label.side=left}{i,v1}
  \fmf{dashes,label=$K_{\ScaGauge{b}}$,label.side=left}{v2,o}
  \fmfdot{v1}
  \fmfv{decor.shape=square,decor.filled=empty,decor.size=8}{v2}
  \fmf{phantom,tension=5}{v2,v3}
  \fmfct{v3}
  \fmf{ghost,left,tension=0.3}{v1,v3}
  \fmf{dashes,right,tension=0.3}{v1,v3}
\end{fmfgraph*}

\end{fmffile}
}


\maketitle

\begin{abstract}
 \noindent 
We compute one-loop and two-loop $\beta$-functions for vacuum
expectation values (VEVs) in gauge theories. In $R_\xi$ gauge the VEVs
renormalize differently from the respective scalar fields. We focus
particularly on the origin and behaviour of this difference and show
that it can be interpreted as the anomalous dimension of a certain
scalar background field, leading to simple direct computation and
qualitative understanding. The results are given for generic as well
as supersymmetric gauge theories. These complement the set of
well-known $\gamma$- and $\beta$-functions of Machacek/Vaughn. As an
application, we compute the $\beta$-functions for VEVs and $\tan
\beta$ in the MSSM, NMSSM, and \ESSM. 
\end{abstract}
\tableofcontents
\clearpage
\section{Introduction}

Local gauge invariance has been established as the underlying
principle of all fundamental interactions. Spontaneously broken gauge
invariance, together with the postulate of a perturbative Higgs
sector, is the basis of the theoretical description of electroweak
interactions in the Standard Model (SM) or extensions like the Minimal
Supersymmetric Standard Model (MSSM). These models have successfully
passed electroweak precision tests \cite{Z-pole} and are in line with the
discovery of a Higgs-like particle at the LHC \cite{ATLAS:2012gk,CMS:2012gu}.

Quantum field theoretical foundations of spontaneously broken gauge
theories like renormalizability, unitarity, and gauge independence of
the S-matrix have been established in
\cite{'tHooft:1971rn,'tHooft:1972ue,Lee:1972fj,Lee:1974zg,Lee:1972yfa}. Later,
BRS invariance and algebraic renormalization have been introduced as
elegant tools \cite{Becchi:1974xu,Becchi:1974md,Becchi:1975nq}. They were used
in the all-order treatments of the renormalizability of the SM
\cite{Kraus:1997bi,Kraus:1998xt,Grassi:1997mc,Grassi:1999nb} and the
MSSM \cite{Hollik:2002mv}.

In the present paper, we focus on the scalar (Higgs) vacuum expectation
values (VEVs) and their renormalization in spontaneously broken gauge
theories. In spite of their obvious central role, the VEVs are no
gauge invariant, physical quantities and therefore less comprehensively
studied. Like Refs.~\cite{Kraus:1997bi,Kraus:1998xt,Hollik:2002mv}, we use the
approach put forward in Ref.~\cite{Kraus:1995jk}, where the VEVs are treated
as background fields, similar to the background field method  
\cite{DeWitt:1967uc,KlubergStern:1974xv,KlubergStern:1975hc,Grassi:1997mc,Grassi:1999nb}.
We describe the renormalization of general gauge theories in this
approach, determine Feynman rules, and compute relevant renormalization
constants. We show that this framework yields several results of
practical and theoretical interest in an elegant way:
\begin{enumerate}
  \item The renormalization transformation for a VEV $v$ can
    generically be written in two equivalent ways,\footnote{%
In Refs.~\cite{Dabelstein:1994hb,Chankowski:1992er}, where the second
form is chosen, our $\delta \bar{v}$ is called $-\delta v$, while our
$\delta v$ is not used.} 
  \begin{align}
  v\to v+\delta v &= \sqrt{Z}\left( v +\delta\bar{v}\right),
  \label{DeltaVDecomposition}
  \end{align}
    where $\sqrt{Z}$ is the field renormalization constant of the 
    respective scalar field and $\delta \bar{v}$ is an extra term,
    which characterizes to what extent the VEV renormalizes
    differently from the field. We
    will show that this extra term has an elegant interpretation in
    terms of the background field and can be computed easily from the
    background field Feynman rules. This will also clarify why the
    $\delta \bar{v}$-term does not appear in theories with only rigid
(global) invariance, and that even in local gauge theories it is
    only required for particular gauges.
\item In many extensions of the SM with several Higgs doublets such as
  the MSSM, the ratio of two VEVs   $\tan\beta=v_u/v_d$ is
  considered. The explicit MSSM calculations of
Refs.~\cite{Dabelstein:1994hb,Chankowski:1992er} have found a
  cancellation, in the notation of Eq.~\eqref{DeltaVDecomposition}
  \begin{align}
  \frac{\delta \bar v_u}{v_u} - \frac{\delta \bar v_d}{v_d} = \mbox{finite}
  \label{eqn_VEV_diff} 
  \end{align}
at the one-loop level. Our approach will make clear that this
cancellation is not general. We will exhibit the origin of the
one-loop cancellation and extend the discussion to the two-loop level
in the MSSM, NMSSM, and \ESSM. The latter two
cases provide examples where Eq.~\eqref{eqn_VEV_diff} is not valid
(see also Refs.~\cite{Yamada:2002nu,Athron:2012pw} for corresponding
results on the $\tan\beta$ renormalization constant).
\item Finally, we compute the renormalization-group $\beta$-functions
  for all VEVs in the general gauge theory and a general supersymmetric gauge
  theory at the one-loop and leading two-loop level. These results
  complement the well-known $\beta$ and $\gamma$ functions of
  Machacek/Vaughn \cite{Machacek:1983tz,Machacek:1983fi,Machacek:1984zw} and
  Martin/Vaughn, Yamada, Jack and
  Jones~\cite{Martin:1993zk,Yamada:1994id,Jack:1994kd} for parameters
  and fields. 
\end{enumerate} 
The outline of the present paper is the following. 
First, we will introduce the generic model together with its
properties and renormalization in
Section~\ref{sec_2_theory}. Sec.~\ref{subsec_3_general_consequences}
discusses the crucial points of our formalism and its equivalence to
the standard approach. The remainder of Sec.~\ref{sec_3_methods} gives an
overview of the necessary one-loop and two-loop computations and
states our main results, the general $\beta$-functions. Finally,
Section~\ref{sec_4_results} 
applies the general results to the MSSM, NMSSM, and \ESSM\ in order to
provide the results for $\tan\beta$ and the VEV $\beta$-functions. 

\section{General Gauge Theory and Scalar Background Fields}
\label{sec_2_theory}
\subsection{Lagrangian}
\label{subsec_2_Lagrangian}
The present paper investigates the renormalization of general, spontaneously
broken gauge theories. Following
Refs.~\cite{Box:2007ss,Luo:2002ti,Machacek:1983tz}, we write the gauge invariant
Lagrangian in terms of real scalar fields $\ScaGauge{a}$ and Weyl
2-spinors $\SpinorDo{p}{\alpha}$ as 
\begin{align}
  \mathcal{L}_{\mathrm{inv}}= &-\frac{1}{4} F_{\mu \nu}^A F^{A \mu \nu} + \frac{1}{2} \left(D_\mu \ScaGauge{} \right)_a \left(D^\mu \ScaGauge{} \right)_a + i \SpinorUp{p}{\alpha} \sigma_{\alpha \dot{\alpha}}^{\mu} \left( D_{\mu}^{\dagger} \AdjSpinorUp{}{\alpha}  \right)_p \nonumber \\
  &- \frac{1}{2!} \MassScaGauge{a}{b} \ScaGauge{a} \ScaGauge{b} - \frac{1}{3!} \TriScaGauge{a}{b}{c} \ScaGauge{a} \ScaGauge{b} \ScaGauge{c} - \frac{1}{4!} \QuadScaGauge{a}{b}{c}{d}  \ScaGauge{a} \ScaGauge{b} \ScaGauge{c} \ScaGauge{d} \\
  &- \frac{1}{2} \left[ \MassSpinor{p}{q} \SpinorUp{p}{\alpha}
\SpinorDo{q}{\alpha} + \mathrm{h.c.} \right]  - \frac{1}{2} \left[
\YukaGauge{p}{q}{a} \SpinorUp{p}{\alpha} \SpinorDo{q}{\alpha} \ScaGauge{a} +
\mathrm{h.c.} \right] . \nonumber
  \label{lagrangian_basic}
\end{align}
Here the covariant derivatives and field strength tensor are defined as 
\begin{subequations}
  \begin{align}
    D_{\mu}\varphi_a&=\left(\delta_{a b} \partial_{\mu} + i g
  \GenGauge{a}{b}{A} \VecGaugeDo{A}{\mu} \right)\ScaGauge{b} , \\
    D_{\mu}\psi_{p \alpha}&=\left(\delta_{p q} \partial_{\mu} + i g
\GenSpiGauge{p}{q}{A} \VecGaugeDo{A}{\mu}\right) \SpinorDo{q}{\alpha} , \\
  F_{\mu \nu}^{A}&= \partial_{\mu} V_{\nu}^{A} - \partial_{\nu} V_{\mu}^{A} - g
f^{A B C}\VecGaugeDo{B}{\mu} \VecGaugeDo{C}{\nu} ,
  \end{align}
\end{subequations}
with antisymmetric, purely imaginary generators $\GenGauge{a}{b}{A}$ for
the scalars; hermitian generators $\GenSpiGauge{p}{q}{A}$ for the
spinors; and structure constants $f^{A B C}$.
The standard procedure in spontaneously broken gauge theories is to shift the
scalar fields by a constant (the ``VEV'')
\begin{align}
 \ScaGauge{a} \rightarrow \ScaGauge{a} + v_{a} ,
  \label{eqn_shift_scalar}
\end{align}
where $v_{a}$ can be adjusted to the minimum of the scalar potential.
After applying the shift \eqref{eqn_shift_scalar}, the Lagrangian
$\mathcal{L}_{\text{inv}}$ is still invariant under both local and global gauge
transformations, if $(\ScaGauge{a} + v_{a})$ are transformed as a whole.

For quantization a gauge fixing is required. In QED and QCD, typical gauge
fixing terms break local gauge invariance but leave global gauge invariance
intact. In contrast, $R_\xi$-gauges, for example, as often used in the
spontaneously broken case, break even global invariance. This breaking is
crucial for the renormalization properties of $v_{a}$. It turns out
that these properties can be studied well by using background fields
\cite{Kraus:1995jk} instead of the shift in Eq.~\eqref{eqn_shift_scalar}. 

In the following we briefly introduce the setup for $R_\xi$ gauges
including background fields.  In
Section~\ref{subsec_3_general_consequences} we will highlight its
crucial points, relate it to the standard procedure, and draw
consequences. We note here only that the background formalism
is a tool that provides additional information but does not alter any
results compared to the standard approach.

We introduce real scalar background fields $(\ScaBackGauge{a}
+ \VevGauge{a})$; $\ScaBackGauge{a}$ is treated as a classical background field,
$\VevGauge{a}$ as a constant. The combination $(\ScaBackGauge{a} +
\VevGauge{a})$, by definition, has the same gauge transformation properties as
$\ScaGauge{a}$. By means of the replacement
\begin{align}
  \ScaGauge{a} \rightarrow \ScaGauge{a}^{\text{eff}}= \ScaGauge{a} +
\ScaBackGauge{a} + \VevGauge{a} 
\label{phieffDefinition}
\end{align}
we introduce a non-trivial ground state as well as the background
field.

Gauge fixing and Slavnov-Taylor identities require BRS transformations. The
background fields transform as a BRS doublet with another background field
$\BrstBackGauge{a}$,
\begin{subequations}
  \begin{align}
 s \ScaBackGauge{a} = \BrstBackGauge{a} , \qquad  \qquad 
s \BrstBackGauge{a} = 0 .
  \end{align}
This implies that the physics content of the theory is unchanged by
including the background fields \cite{Piguet:1995er,Brandt1989263}. The BRS
transformations of the scalar fields read
  \begin{align}
 s \ScaGauge{a} = - ig \GenGauge{a}{b}{A} \GhoGauge{A}
\ScaGauge{b}^{\text{eff}} -\BrstBackGauge{a} ,
  \label{eqn_BRS_phi}
  \end{align}
such that $\ScaGauge{a}^{\text{eff}}$ transforms homogeneously
  \begin{align}
  s \ScaGauge{a}^{\text{eff}} = - ig \GenGauge{a}{b}{A} \GhoGauge{A}
\ScaGauge{b}^{\text{eff}} .
  \end{align}
All other BRS transformations are standard
\cite{Becchi:1974xu,Becchi:1974md,Becchi:1975nq} and read in our notation
  \begin{align}
    s \VecGaugeDo{A}{\mu} &= \partial_{\mu} \GhoGauge{A} - g f^{A B C}
\VecGaugeDo{B}{\mu} \GhoGauge{C} , \\
    s \SpinorDo{p}{\alpha}  &= - i g \GhoGauge{A} \GenSpiGauge{p}{q}{A}
 \SpinorDo{q}{\alpha} 	, \\
    s \GhoGauge{A} &= \frac{1}{2} g f^{A B C} \GhoGauge{B} \GhoGauge{C},  \\
    s \AntiGhoGauge{A} &= \NaLauGauge{A},  \qquad \qquad  s
\NaLauGauge{A} = 0 .  
  \end{align}
\end{subequations}
Herein, $\GhoGauge{A}$, $\AntiGhoGauge{A}$, and $\NaLauGauge{A}$ denote the
Faddeev-Popov ghost, Faddeev-Popov antighost, and the Nakanishi-Lautrup
auxiliary field, respectively.

If not stated otherwise, we will use $R_\xi$ gauge fixing with gauge fixing
function
\begin{align}
 F^A = \partial^{\mu} \VecGaugeDo{A}{\mu}  +i g \xi \xi' \left(\ScaBackGauge{} +
\VevGauge{} \right)_a \GenGauge{a}{b}{A} \ScaGauge{b}.
\end{align}
We will always set $\xi'=1$, but keep it as a variable because it is
renormalized. The full gauge fixing and ghost Lagrangian is then given by
evaluating
\begin{subequations}
  \begin{align}
    \mathcal{L}_{\text{fix, gh}}&= s \left[ \AntiGhoGauge{A} \left( F^A +
\frac{\xi}{2} \NaLauGauge{A} \right) \right],
    \label{eqn_def_fix-gh}
  \end{align}  
which yields after elimination of $\NaLauGauge{A}$
  \begin{align}
   \mathcal{L}_{\text{fix, gh}} = &-\frac{1}{2 \xi} (\partial^{\mu}
\VecGaugeDo{A}{\mu}) (\partial^{\nu} \VecGaugeDo{A}{\nu}) - \AntiGhoGauge{A}
\Box \GhoGauge{A} - g f^{A B C} (\partial^{\mu}
\AntiGhoGauge{A})\VecGaugeDo{B}{\mu} \GhoGauge{C} \label{eqn_lagragian_gf} \\
    &-i g \xi' (\partial^{\mu} \VecGaugeDo{A}{\mu}) (\ScaBackGauge{} +
\VevGauge{}
)_a \GenGauge{a}{b}{A} \ScaGauge{b} - i g \xi \xi' \AntiGhoGauge{A}
\BrstBackGauge{a} \GenGauge{a}{b}{A} (\ScaGauge{} + \ScaBackGauge{} +
\VevGauge{} )_b \nonumber \\
    & -g^2 \xi \xi' \AntiGhoGauge{A} \GhoGauge{B} (\ScaBackGauge{}
+\VevGauge{})_a \GenGauge{a}{b}{A} \GenGauge{b}{c}{B} (\ScaGauge{} +
\ScaBackGauge{} + \VevGauge{} )_c \nonumber \\
    &+ \frac{1}{2} g^2 \xi \xi'^2 (\ScaBackGauge{} +\VevGauge{})_a
\GenGauge{a}{b}{A}\ScaGauge{b} (\ScaBackGauge{} +\VevGauge{})_c
\GenGauge{c}{d}{A}\ScaGauge{d} . \nonumber
  \end{align}
\end{subequations}
This modified $R_\xi$ gauge fixing preserves the rigid invariance
due to the background fields. Finally, the study of renormalization
is streamlined by introducing sources $K$ for the non-linear BRS
transformations 
\begin{align}
  \mathcal{L}_{\mathrm{ext}} = K_{\ScaGauge{a}} \, s
\ScaGauge{a} + K_{\VecGaugeDo{A}{\mu}} \, s
\VecGaugeDo{A}{\mu} +  K_{\GhoGauge{A}} \, s \GhoGauge{A} + \left[
K_{\Spinor{p}} \, s \Spinor{p} + \mathrm{h.c.} \right].
  \label{eqn_Lagrangian_ext}
\end{align}
In summary, the total Lagrange density is the sum
of all discussed parts,
\begin{align}
  \mathcal{L}_{\mathrm{tot}}= \mathcal{L}_{\mathrm{inv}}|_{\varphi \rightarrow
\varphi^{\text{eff}}} + \mathcal{L}_{\text{fix, gh}} +
\mathcal{L}_{\mathrm{ext}} .
  \label{lagrangian_total}
\end{align}

\subsection{Renormalization}
\label{subsec_2_renormalization}
Renormalization proceeds essentially as in the case without background fields.%
\footnote{In particular cases, such as the SM and the MSSM in
  Refs.~\cite{Kraus:1997bi,Kraus:1998xt,Hollik:2002mv}, the background 
fields were essential in order to control the $U(1)_{\text{em}}$ Ward identity.
In the MSSM an even more complicated background field structure has been used
to allow for on-shell renormalization conditions separating unphysical from
physical degrees of freedom. Here, however, we are concerned with the generic
situation and minimal subtraction, where the background fields are optional.}
In the usual case the divergence structure is controlled by identities such as
(1) the Slavnov-Taylor identity, expressing BRS invariance on the level of Green
functions; (2) the so-called gauge condition, fixing the
$\NaLauGauge{A}$-dependence, i.e.\ the $R_\xi$ gauge fixing term. All divergences
can be absorbed by a multiplicative renormalization transformation of fields and
parameters. The required form is obtained from the general classical solution of
the Slavnov-Taylor identity and the gauge condition
\cite{Becchi:1974xu,Becchi:1974md,Becchi:1975nq}.

Furthermore, the behaviour of $\mathcal{L}_{\text{fix, gh}}$ under rigid gauge
transformations is crucial for the renormalization of the shift $v$. By
construction, $\mathcal{L}_{\text{fix, gh}}$ necessarily breaks local
gauge invariance. Nevertheless, some gauges like Landau gauge ($\xi=0$)
respect rigid gauge invariance. In such cases the
corresponding rigid Ward identity leads to the combined renormalization of
$\ScaGauge{} + v$, i.e.\ the additional counterterm $\delta \bar{v}$ in
Eq.\ \eqref{DeltaVDecomposition} is
forbidden by symmetry. If, on the other hand, $\mathcal{L}_{\text{fix, gh}}$
breaks rigid gauge invariance then no symmetry forbids $\delta \bar{v}$,
implying that $\delta \bar{v}$ will in general be necessary and divergent.
Without background fields we have no control of $\delta \bar{v}$.

With background fields, again divergences are controlled by the Slavnov-Taylor
identity and gauge-condition --- but, in addition, we have one more identity at
our disposal: the rigid Ward identity expressing rigid gauge invariance in the
presence of $\ScaBackGauge{}$ and $\BrstBackGauge{} $. Since it holds
in the modified $R_\xi$ gauge \eqref{eqn_lagragian_gf} we gain further
information.

Going through the standard steps and taking the additional Ward identity into
account we obtain the most general structure of the divergences. The resulting
renormalization transformations required to absorb all divergences can be
summarized as follows:
\begin{enumerate}
  \item Parameter renormalization: All parameters $p \in \{
g,\xi,\xi',\MassScaGauge{a}{b} ,\TriScaGauge{a}{b}{c}, \QuadScaGauge{a}{b}{c}{d}
, \MassSpinor{p}{q}, \YukaGauge{p}{q}{a} \}$ renormalize as $p \rightarrow p +
\delta p$. The shift $\VevGauge{}$ does not.
  \item Field renormalization: All fields transform multiplicatively with
appropriate $\sqrt{Z}$ factors. In particular, the renormalization
transformations of interest are  
  \begin{subequations}
    \begin{align}
    \ScaGauge{a} &\rightarrow \sqrt{Z}_{a b} \ScaGauge{b}, \label{eqn_RT_scalar}
\\
    ( \ScaBackGauge{} + \VevGauge{} )_a &\rightarrow \sqrt{Z}_{a b}
\sqrt{\hat{Z}}_{b c} ( \ScaBackGauge{} + \VevGauge{} )_c,
\label{eqn_RT_background} \\
    \BrstBackGauge{a} &\rightarrow \sqrt{Z}_{a b} \sqrt{\hat{Z}}_{b c}
\BrstBackGauge{c}. \label{eqn_RT_BRS_background}
    \end{align}
  \end{subequations}
  Eq.~\eqref{eqn_RT_background} is a consequence of the unbroken rigid
  gauge invariance, and Eq.~\eqref{eqn_RT_BRS_background} stems from
  the fact that $\BrstBackGauge{a}$ is the BRS transformation of
  $\ScaBackGauge{a}$. The BRS sources $K$ transform with inverse
  $\sqrt{Z}$ factors, in particular
  \begin{align}
    K_{\ScaGauge{a}} \rightarrow \left(\sqrt{Z}^{-1} \right)_{ba} 
K_{\ScaGauge{b}}.
    \label{eqn_cond_RT_BRS_source}
  \end{align}
\end{enumerate}
The above mentioned relations prohibit an additional $\VevGauge{}$ counterterm
and simplify the counterterm structure of the two-point function
between $K_{\ScaGauge{}}$ and $\BrstBackGauge{}$.

Before moving on to the next section, we impose a further constraint on the
theories we consider. We require that the theories possess some additional
symmetry, at least at the dimension 4 level, that enforces the field
renormalization for the real scalar fields to be diagonal
  \begin{align}
    \sqrt{Z}_{a b} \rightarrow \sqrt{Z}(a) \delta_{a b} \qquad \text{and} \qquad
\sqrt{\hat{Z}}_{a b} \rightarrow \sqrt{\hat{Z}}(a) \delta_{a b}.
\label{DiagonalZFactors}
  \end{align}
  As an example, in the MSSM and NMSSM this symmetry is realized by the
so-called Peccei-Quinn (PQ) symmetry \cite{Peccei:1977hh} (only softly broken
by the so-called $\mu$-term), whereas in the \ESSM\ the
additional $U(1)_N$ gauge group plays this role.%
\footnote{In these models these symmetries and Eq.\
\eqref{DiagonalZFactors} are the reason why no off-diagonal kinetic
counterterms like $\delta Z_{H_u H_d} \epsilon_{ij} (D^\mu H_u)_i (D_\mu
H_d)_j$ (where $i$, $j$ are $SU(2)$ indices and $\epsilon_{ij}$ is
antisymmetric) are necessary, in spite of their gauge invariance. Note
that Eq.\ \eqref{DiagonalZFactors} does not forbid introducing
additional, finite off-diagonal $Z$-factors e.g.\ for defining mass
eigenstate fields as discussed in detail e.g.\ in Ref.\
\cite{Hollik:2002mv}.} 

\section{General Results}
\label{sec_3_methods}
\subsection{Equivalence and General Consequences for \texorpdfstring{$\delta v$}{delta v}}
\label{subsec_3_general_consequences}
The purpose of this section is to emphasize and exploit the features of
the approach explained above in Section~\ref{sec_2_theory} and explain
its equivalence to the standard approach.

In the standard approach, without background fields, the most generic
renormalization transformation of the scalar fields with shifts reads
as
\begin{subequations}
  \begin{align}
 \ScaGauge{a} + v_{a} \rightarrow \sqrt{Z}_{a b} \left( \ScaGauge{b} + v_{b}
+ \delta \bar{v}_{b} \right),
  \label{eqn_RT_phi_standard}
  \end{align}
whereas we have, from Eqs.\
\eqref{phieffDefinition}, \eqref{eqn_RT_scalar} and \eqref{eqn_RT_background},
  \begin{align}
  \ScaGauge{a}^{\text{eff}} \rightarrow \sqrt{Z}_{a b} \left(\ScaGauge{b} +
 \sqrt{\hat{Z}}_{b c} ( \ScaBackGauge{} + \VevGauge{} )_c \right).
     \label{eqn_RT_phi_eff}
  \end{align}
\end{subequations}
For the calculation of Green functions the background field has to be
set to zero, $\ScaBackGauge{}=0$. 
Hence, the comparison of Eq.~\eqref{eqn_RT_phi_standard} and Eq.~\eqref{eqn_RT_phi_eff}
yields the following identification between the two formalisms:
\begin{align}
  v_{a} + \delta \bar{v}_{a} = \sqrt{\hat{Z}}_{a b} \VevGauge{b},
\label{Correspondence1}
\end{align}
and equivalently
\begin{subequations}
\begin{align}
 v_a &= \VevGauge{a},\\
 \delta \bar{v}_a &= \left(\sqrt{\hat{Z}} - 1 \right)_{a b}
 \VevGauge{b} = \frac{1}{2} \delta \hat{Z}_{a b} \VevGauge{b} +
 \mathcal{O}(\hbar²),\label{eq:general-vbar}\\
 \delta v_a &\overset{\eqref{DeltaVDecomposition}}{=}
 \left( \sqrt{Z} \sqrt{\hat{Z}} - 1 \right)_{ab} \VevGauge{b}
= \frac{1}{2} \left( \delta Z + \delta\hat{Z} \right)_{ab} \VevGauge{b} +
 \mathcal{O}(\hbar²). \label{eq:v-decomposition-1loop}
\end{align}
\end{subequations}
The advantage of this method is that one possesses more information about the
renormalization properties of the shift $\VevGauge{}$. First, the field
renormalization $\hat{Z}$ is a dimension zero quantity which is at most
logarithmic divergent. Second, the shift-counterterm $\delta \bar{v}_a$ is
linear in
the shift itself and otherwise determined by $\hat{Z}$. Third,
$\hat{Z}$ is the field renormalization of the background field, meaning
that it appears not only in Eq.~\eqref{eqn_RT_phi_eff} but also in other Green
functions and, thus, can be directly evaluated by certain diagrams.

To elaborate on the last point, consider the computation of $\sqrt{\hat{Z}}$. It
turns out that $\sqrt{\hat{Z}}$ is the only renormalization constant appearing
in the two point function $\Gamma_{\BrstBackGauge{a},K_{\ScaGauge{b}}}$, 
\begin{align}
\mathcal{L}_{\mathrm{ext}} = - K_{\ScaGauge{a}} \BrstBackGauge{a} + \cdots
\stackrel{\text{RT}}{\rightarrow}  - K_{\ScaGauge{a}} \sqrt{\hat{Z}}_{ab}
\BrstBackGauge{b} + \cdots ,
\end{align}
leading to the Feynman rules in
Fig.~\ref{fig_Feynman_rules_hat_Z_one-loop} (wherein the cross denotes the
one-loop counterterm $\delta \hat{Z}$).
\begin{figure}
  \centering
  \subfigure[$\Gamma_{\ScaGauge{b},\GhoGauge{A}, K_{\ScaGauge{a}}}$]{%
    $\fmfvcenter{ThreePtBrstSource1Sca1Gh1} \, = \, g \GenGauge{a}{b}{A}$\,,
    \label{fig_BRS-source_ghost_scalar_vertex}
  }
  \quad
  \subfigure[$\Gamma_{\ScaGauge{b}, \BrstBackGauge{a}, \AntiGhoGauge{A}}$]{%
    $\fmfvcenter{ThreePtBrstBack1Sca1AntiGh1} \, = \, \xi \xi' g
\GenGauge{a}{b}{A}$\,,
    \label{fig_BRS_background-field_anti-ghost_scalar_vertex}
  }%
  \quad
  \subfigure[$\Gamma_{\BrstBackGauge{a},K_{\ScaGauge{b}}}$]{%
    $\fmfvcenter{TwoPtCTQandBRST} \, = \, - \frac{i}{2} \delta \hat{Z}_{b
a}^{(1)}$
    \label{fig_BRS-source_BRS_background-field_CT}
  }%
  \caption{Feynman rules for $\sqrt{\hat{Z}}$ determination.}
  \label{fig_Feynman_rules_hat_Z_one-loop}
\end{figure}
Hence, $\sqrt{\hat{Z}}$ can be directly determined from the divergence to that
two point function. Of course, $\Gamma_{\BrstBackGauge{a},K_{\ScaGauge{b}}}$ is
an unphysical Green function, which highlights its role as technical tool.
However, the point is that very few Feynman diagrams with well localized origin
contribute to it. The only coupling of the field $\BrstBackGauge{}$ to
propagating fields is the term $\sim \AntiGhoGauge{} \BrstBackGauge{} 
\ScaGauge{}$ in $\mathcal{L}_{\text{fix, gh}}$ (see
Fig.~\ref{fig_BRS-source_ghost_scalar_vertex}), which stems from BRS
invariance. Its coefficient must be the same as the one of the $(\ScaBackGauge{}
+ \VevGauge{}) \ScaGauge{}$-term in the gauge fixing function $F^A$. Similarly,
the term $\sim K_{\ScaGauge{}} \GhoGauge{} \ScaGauge{}$ is determined by
inserting the BRS transformation $s \ScaGauge{}$ in $\mathcal{L}_{\text{ext}}$,
see Eq.~\eqref{eqn_Lagrangian_ext} and
Fig.~\ref{fig_BRS_background-field_anti-ghost_scalar_vertex}. Most
important, all the mentioned Feynman rules are proportional to the gauge
coupling $g$, hence, $\sqrt{\hat{Z}}$ is (at one-loop level) proportional to two
powers of gauge couplings with coefficients fixed by BRS invariance. In
contrast, all dimensionless couplings can contribute to the field
renormalization constant $\sqrt{Z}$.

In summary, the method of background fields together with BRS and
rigid invariance provides the following features:  the VEV counterterm
$\delta  \bar{v}_a$ from Eq.\ \eqref{DeltaVDecomposition} is given by
$\sqrt{\hat{Z}}$, 
see Eq.\ \eqref{eq:general-vbar}. Hence it is
proportional to the VEV itself, and its divergence is given by a
dimension zero quantity and at most logarithmic divergent. The
relevant field renormalization 
$\hat{Z}$ can be determined by a single two point function to which few,
specific Feynman rules contribute.
 
\subsection{Field Renormalization of Scalar Fields and Background
Fields}
This section will serve for the one- and leading two-loop renormalization
computations. All calculations are performed in $R_\xi$ gauge and the
\MSbar /\DRed\ scheme.%
\footnote{\DRed\ denotes modified minimal subtraction in
  regularization by dimensional reduction. There is no difference
  between the two schemes in any of the calculations carried out here.}
 
\subsubsection{One-loop}
The one-loop results for the divergent parts of $Z$ and $\hat{Z}$ will be
provided in this section together with a brief overview of the relevant
diagrams. 

The scalar field renormalization constant $\sqrt{Z}$ is given by the
derivative of the scalar self-energy with respect to external momentum squared.
At one-loop, there are ten (one-particle irreducible) diagrams and
renormalization works as usual by requiring 
  \begin{align}
  \fmfvcenter{ScaSelfBlob}\quad + \quad \fmfvcenter{ScaSelfCT}  = \text{finite}.
  \end{align}
According to Refs.~\cite{Machacek:1983tz,Luo:2002ti}, the results are
well-known; in our notation they are given by
\begin{subequations}
  \begin{align}
  \delta_{a b} \delta Z^{(1)}(a)&= \frac{1}{(4
    \pi)^2} \left[ g^2 \left(3- \xi\right) \CasiGauge{a}{b} -
    \YukaInvGauge{a}{b} \right] \cdot \frac{1}{\epsilon}
      \label{eqn_Z_CT} \\
    &= \gamma_{ab}^{(1)}(\mathrm{S}) \cdot \frac{1}{\epsilon},
  \end{align}
\end{subequations}
where $D=4 - 2 \epsilon$ in dimensional regularization,
$\gamma_{ab}^{(1)}(\mathrm{S})$ is the one-loop anomalous dimension, and we
introduced the following group invariants (for the scalar representation $S$) 
\begin{align}
 \CasiGauge{a}{b} = \GenGauge{a}{c}{A} \GenGauge{c}{b}{A}, \qquad  \qquad
  \YukaInvGauge{a}{b}= \frac{1}{2} \left( \YukaGauge{p}{q}{a}
\YukaGaugeConj{p}{q}{b} + \YukaGaugeConj{p}{q}{a} \YukaGauge{p}{q}{b} \right).
\end{align}

As described above, the new renormalization constant $\sqrt{\hat{Z}}$ can be
determined easily because it appears as the counterterm Feynman rule for
$\Gamma_{\BrstBackGauge{a},K_{\ScaGauge{b}}}$. There exists only one diagram in
one-loop order. Hence, requiring 
\begin{align}
 \fmfvcenter{TwoPtSca1Gh1}  \quad + \quad \fmfvcenter{TwoPtCTQandBRST} =
\text{finite}
\end{align}
leads to the result
  \begin{align}
 \delta_{a b} \delta \hat{Z}^{(1)}(a) = \frac{1}{(4 \pi)^2} 2 g^2 \xi \xi'
\CasiGauge{a}{b} \cdot \frac{1}{\epsilon} .
\label{eqn_Z_hat_CT}
  \end{align}

\subsubsection{Two-loop}
At the two-loop level, a similar computation of the two-loop self energy
determines $\delta Z^{(2)}$.  It is convenient to express the two-loop
renormalization parameter in terms of the anomalous dimension
$\gamma(\mathrm{S})$ and $\beta$ functions
\begin{align}
 \frac{1}{2} \delta Z_{ab}^{(2)} &= \frac{1}{4} \gamma_{ab}^{(2)} (\mathrm{S})
\cdot \frac{1}{\epsilon} + \frac{1}{8} \left[ \gamma_{ac}^{(1)} (\mathrm{S})
\gamma_{cb}^{(1)} (\mathrm{S}) + \sum_x \beta^{(1)}(x) \left( \partial_x
\gamma_{ab}^{(1)} (\mathrm{S}) \right) \right] \cdot \frac{1}{\epsilon^2},
\end{align}
wherein $x\in \{ g,\xi,\xi',Y_{pq}^{a},Y_{pq}^{*a}\}$. The calculations have
been performed in Refs.~\cite{Machacek:1983tz,Luo:2002ti} and the
anomalous dimension for the scalar fields (in DREG) is given as
\begin{align}
 \gamma_{a b}^{(2)} (\mathrm{S}) &= \frac{1}{(4 \pi )^4} \Bigg\{ g^4
\CasiGauge{a}{b}\left[\left(\frac{35}{3} - 2 \xi -\frac{1}{4} \xi^2 \right) C_2
(\mathrm{G}) - \frac{10}{6} S_2(\mathrm{F}) - \frac{11}{12} S_2(\mathrm{S})
\right] \\
 &\phantom{= \frac{1}{(4 \pi )^4} \Bigg\{}
 -\frac{3}{2}g^4\CasiGauge{a}{c}\CasiGauge{c}{b} +\frac{3}{2}
H_{ab}^{2}(\mathrm{S}) + \bar{H}_{ab}^{2}(\mathrm{S}) - \frac{10}{2} g^2
Y_{ab}^{2 F}(\mathrm{S}) - \frac{1}{2} \Lambda_{ab}^{2}(\mathrm{S}) \Bigg\} ,
\nonumber
\end{align}
where the group invariants $C_2(\mathrm{G})$, $S_2(\mathrm{F})$,
$S_2(\mathrm{S})$, $\Lambda_{ab}^{2}(\mathrm{S})$, $H_{ab}^{2}(\mathrm{S})$,
$\bar{H}_{ab}^{2}(\mathrm{S})$, and $Y_{ab}^{2 F}(\mathrm{S})$ are defined as in
Ref.~\cite{Luo:2002ti}.

\begin{figure}[htpb!]
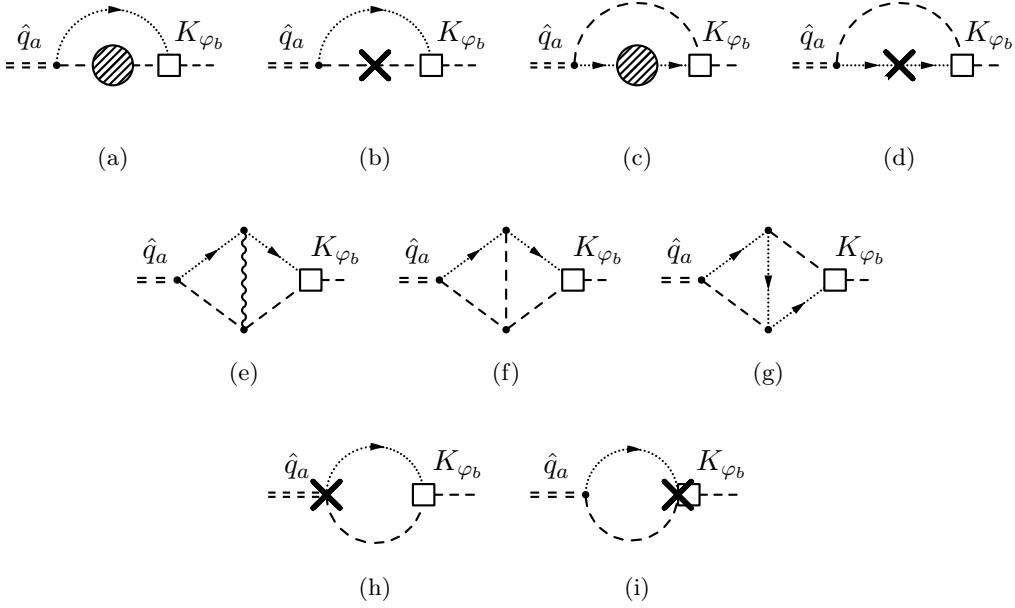

  \centering
  \subfigure[]{%
    $\fmfvcenter{TwoPtBrstBackTwoLoopScaSelf}$
    \label{fig_insertion_scalar_selfenergy}
  }
  \quad
  \subfigure[]{%
    $\fmfvcenter{TwoPtBrstBackTwoLoopScaSelfCT}$
    \label{fig_insertion_scalar_counterterm}
  }
  \quad
  \subfigure[]{%
    $\fmfvcenter{TwoPtBrstBackTwoLoopGhoSelf}$
    \label{fig_insertion_ghost_selfenergy}
  }
  \quad
  \subfigure[]{%
    $\fmfvcenter{TwoPtBrstBackTwoLoopGhoSelfCT}$
    \label{fig_insertion_ghost_counterterm}
  }

  \subfigure[]{%
    $\fmfvcenter{TwoPtBrstBackTwoLoopVecExchange}$
    \label{fig_exchange_vector}
  }
  \quad
  \subfigure[]{%
    $\fmfvcenter{TwoPtBrstBackTwoLoopScaExchange}$
    \label{fig_exchange_scalar}
  }
  \quad
  \subfigure[]{%
    $\fmfvcenter{TwoPtBrstBackTwoLoopGhoExchange}$
    \label{fig_exchange_ghost}
  }

  \subfigure[]{%
    $\fmfvcenter{TwoPtBrstBackTwoLoopCTqVertex}$
    \label{fig_vertex_CT_L_fig-gh}
  }
  \quad
  \subfigure[]{%
    $\fmfvcenter{TwoPtBrstBackTwoLoopCTbrstVertex}$
    \label{fig_vertex_CT_L_ext}
  }
  \caption{Feynman diagrams for full two-loop computation of $\delta \hat{Z}$.
Note that Fig.~\ref{fig_exchange_scalar} and \ref{fig_exchange_ghost} are
power-counting finite. Furthermore, Fig.~\ref{fig_vertex_CT_L_fig-gh} is zero
as the contained vertex counterterm vanishes due to non-renormalization of
$\mathcal{L}_{\text{fix, gh}}$.}
  \label{fig_two-loop_hat_Z}
\end{figure}
Likewise, the two-loop value of $\delta \hat{Z}^{(2)}$ is determined by the
two-loop part of $\Gamma_{\BrstBackGauge{a},K_{\ScaGauge{b}}}$. The relevant
diagrams for the full two-loop corrections to
$\Gamma_{\BrstBackGauge{a},K_{\ScaGauge{b}}}$ are shown in
Fig.~\ref{fig_two-loop_hat_Z}. Here,
Fig.~\ref{fig_insertion_scalar_selfenergy}--\ref{fig_insertion_ghost_counterterm}
contain all insertion of one-loop 
self-energies (shaded circles) and corresponding one-loop counterterms
(crosses). On the other hand,
Fig.~\ref{fig_exchange_vector}--\ref{fig_exchange_ghost} display the
exchange of intermediate fields and 
Fig.~\ref{fig_vertex_CT_L_fig-gh}--\ref{fig_vertex_CT_L_ext} provide
the one-loop vertex counterterms. The inspection of the diagrams leads
to the conclusion that each contribution is either proportional to $g^4$ or $g^2
\YukaGauge{}{}{} {\YukaGauge{}{}{}}^\dagger$. Hence, terms with a different
coupling structure cannot enter $\delta \hat{Z}^{(2)}$. In contrast,
 $\delta Z^{(2)}$ contains terms proportional to $\lambda^4 \leftrightarrow
\Lambda_{ab}^{2}(\mathrm{S})$ or $(\YukaGauge{}{}{}
{\YukaGauge{}{}{}}^\dagger)^2 \leftrightarrow H_{ab}^{2}(\mathrm{S}),
\bar{H}_{ab}^{2}(\mathrm{S})$.

For the purpose of the present paper we restrict ourselves to the
Yukawa-enhanced contributions of order $g^2 Y Y^\dagger $. They are obtained
from requiring
\begin{align}
 \fmfvcenter{TwoPt2LSca1Gh1Spi2} \quad + \quad
\fmfvcenter{TwoPtCT2LSca1Gh1Spi2} \quad + \quad \fmfvcenter{TwoPt2LCTQandBRST}
\quad = \text{finite},
\end{align}
where the crosses denote counterterm contributions from the indicated
renormalization constants. The result reads as 
\begin{align}
\delta \hat{Z}^{(2)}(a)|^{g^2 \YukaInvGauge{}{}}  \, \delta_{a
b} = \frac{1}{( 4 \pi)^4} g^2 \xi \xi'
\GenGauge{a}{c}{A} \YukaInvGauge{c}{d} \GenGauge{d}{b}{A} \left[
\frac{1}{\epsilon^2} - \frac{1}{\epsilon} \right].
\end{align}

\subsection{Results for \texorpdfstring{$\delta v$}{delta v} in Generic Models}
The decomposition of $\delta v_a$ in
Eq.~\eqref{eq:v-decomposition-1loop} leads to an equivalent
decomposition of the $\beta$-function for $v_a$
\begin{align}
 \beta(v_{a})= \mu \partial_{\mu} v_{a}  = \left[ \gamma_{a
b}(\mathrm{S}) + \hat{\gamma}_{a b}(\mathrm{S}) \right] v_{b} ,
  \label{eqn_vev_beta_general}
\end{align}
where $\mu$ is the \MSbar/\DRed\ renormalization scale,
$\gamma(\mathrm{S})$ is the anomalous dimension of the scalar field,
\begin{align}
  \gamma_{a b}(\mathrm{S}) = \left( \mu \partial_{\mu} \sqrt{Z}_{a c}^{-1}
\right) \sqrt{Z}_{cb} ,
\end{align}
 and $\hat{\gamma}(\mathrm{S})$ the analogous quantity for
 $\sqrt{\hat{Z}}$. 

Hence, our main results for VEV renormalization constants and $\beta$
and $\gamma$ functions in the \MSbar\ and \DRed\ schemes can be
summarized as
  \begin{subequations}
    \begin{align}
  \delta v_{a}^{(1)} &= \frac{1}{(4 \pi)^2} \left[ g^2 \left(\frac{3-\xi}{2} +
\xi \xi'\right) \CasiGauge{a}{b} - \frac{1}{2} \YukaInvGauge{a}{b} \right] \,
v_{b} \cdot \frac{1}{\epsilon} , \label{eqn_vev_CT} \\
  \beta^{(1)}(v_{a}) &= \frac{1}{(4 \pi)^2} \left[ g^2 \left(3-\xi + 2
\xi \xi'\right) \CasiGauge{a}{b} - \YukaInvGauge{a}{b} \right] v_{b},
\label{eqn_vev_beta_one-loop_generic} \\
  \gamma_{a b}^{(1)}(\mathrm{S}) &= \frac{1}{(4 \pi)^2} \left[ g^2
    \left(3-\xi \right) \CasiGauge{a}{b} - \YukaInvGauge{a}{b} \right], \\
   \hat{\gamma}_{a b}^{(1)}(\mathrm{S}) &= \frac{1}{(4 \pi)^2} 2 g^2 \xi \xi'
\CasiGauge{a}{b},
    \end{align}
      \label{eqn_general_one-loop_results}%
\end{subequations}
at one-loop level, and
  \begin{subequations}
   \begin{align}
    \beta^{(2)}(v_{a}) &= \gamma_{a b}^{(2)}(\mathrm{S}) v_{b} - \frac{1}{( 4
\pi)^4} 2 g^2 \xi \xi' \GenGauge{a}{c}{A} \YukaInvGauge{c}{d}
\GenGauge{d}{b}{A}  v_{b} + \mathcal{R}_{a b} v_b, \\
    \hat{\gamma}_{a b}^{(2)}(\mathrm{S}) &=  -\frac{1}{( 4\pi)^4} 2 g^2 \xi \xi'
\GenGauge{a}{c}{A} \YukaInvGauge{c}{d} \GenGauge{d}{b}{A}  + \mathcal{R}_{a b}
   \end{align}
    \label{eqn_general_two-loop_results}%
  \end{subequations}
at two-loop level. Here $\mathcal{R}_{a b}$ contains all $1/\epsilon$-pole 
contributions from $\sqrt{\hat{Z}} $ proportional to $g^4$. We remind
the reader that $\xi'=1$, but it is kept as a variable in order to
visualize the origin of the different terms.

\subsection{Results for \texorpdfstring{$\delta v$}{delta v} in
General SUSY Models}
\label{subsec_3_SUSY}

Our results can now be specialized to general supersymmetric (SUSY)
models with spontaneous gauge symmetry breaking.  For supersymmetric
models with or without soft SUSY breaking the results will be the
same, since the general results
\eqref{eqn_general_one-loop_results}--\eqref{eqn_general_two-loop_results}
depend only on dimensionless couplings.  We do not need to specify
soft SUSY breaking terms explicitly, even though the later examples
will be realistic models with softly broken SUSY.

The application to
general supersymmetric models requires one to take gauginos $\lambda^A$ into
account. The generic supersymmetric Lagrangian in Wess-Zumino
gauge with non-abelian gauge interactions contains the standard kinetic terms
for complex scalar fields $\phi_{a}$, their SUSY Weyl-spinor partners $\psi_a$,
gauge fields $\VecGaugeDo{A}{}$, and their SUSY Weyl-spinor partner $\lambda^A$.
Besides a scalar potential with $\phi^n$ interactions for $n \in \{1,2,3,4\}$,
the Lagrangian contains two sets of Yukawa-type interactions
\begin{align}
 \mathcal{L}_{\mathrm{SUSY}} = -  \frac{1}{2}\left[ \psi_p^{\alpha}
\psi_{q \alpha} \mathcal{W}_{p q}  + \text{h.c.} \right] - \sqrt{2} g
\left[\bar{\lambda}_{\dot{\alpha}}^{A } \bar{\psi}_{p}^{\dot{\alpha}}
\GenGauge{p}{q}{A} \phi_q + \text{h.c.}  \right]  + \cdots \,,
\end{align}
where $\mathcal{W}_{p q}$ denotes derivatives of the superpotential
$\mathcal{W}$. Following Ref.~\cite{Martin:1993zk} the superpotential
(formulated in chiral superfields $ \Phi$) is given by
\begin{align}
\mathcal{W} = \frac{1}{3!} Y^{pqr} \Phi_p \Phi_q \Phi_r + \frac{1}{2!} \mu^{p
q} \Phi_p \Phi_q + L^{p} \Phi_p.
\end{align}
Gaugino couplings are special in the sense that their form is given by the gauge
coupling strength $g$ times group generator. Hence, those Yukawa-type couplings
contribute to the $g^2$ part of $\gamma$. The results can be expressed in
terms of $\gamma$ and $\beta$ for the complex scalar fields
$\phi_{a}$. We obtain
\begin{subequations}
  \label{eqn_vev_beta_one-loop_susy}
  \begin{align}
  \gamma_{aa}^{(1)}(\mathrm{S}) \Big|_{\text{SUSY}} &= \frac{1}{(4 \pi)^2}
\left[ g^2
    \left(1-\xi \right) \CasiGauge{a}{a} - \frac{1}{2}\YukaInvGauge{a}{a}
\right],  \\ 
   \hat{\gamma}_{aa}^{(1)}(\mathrm{S}) \Big|_{\text{SUSY}} &= \frac{1}{(4
\pi)^2} 2 g^2 \xi \xi' \CasiGauge{a}{a}, \\
  \beta^{(1)}(v_{a}) \Big|_{\text{SUSY}} &= \frac{1}{(4 \pi)^2} \left[ g^2
\left(1-\xi + 2 \xi \xi'\right) \CasiGauge{a}{a} -  \frac{1}{2}
\YukaInvGauge{a}{a} \right] v_{a} ,
  \end{align}
\end{subequations}  
with the standard convention $\YukaInvGauge{a}{b}= (Y^{pqa} Y^{* pqb} +
Y^{* pqa} Y^{pqb})/2$. Note two changes here compared to the general
results in Eqs.\
\eqref{eqn_general_one-loop_results}: first, the $g^2$ pre-factor has 
changed from $(3-\xi)$ to $(1-\xi)$ due to the Gaugino couplings. Second, the
overall normalization of the Yukawa couplings is different compared to the
Eq.~\eqref{lagrangian_basic} and gives rise to the factor $1/2$ in front of
$\YukaInvGauge{a}{b}$. 
Furthermore, we do not observe a change of the one-loop $\hat{\gamma}$ in a
generic SUSY theory because Yukawa couplings do not contribute to it.

As a remark, we stress again that $\gamma$ is the anomalous dimension
of the component field $\phi$ in Wess-Zumino gauge, which is different
from the corresponding quantity for a superfield $\Phi$ in a
supersymmetric gauge (see Ref.\ \cite{Capper:1984zy} for an explicit
one-loop comparison). The latter can be found in
Ref.~\cite{Martin:1993zk}, but are not relevant in this analysis.

The same considerations are valid at two-loop level and lead to a
corresponding change in $\gamma^{(2)}(\mathrm{S})$ as well as in
$\hat{\gamma}^{(2)}(\mathrm{S})$. However, the $g^2 Y Y^{\dagger}$ part
of $\hat{\gamma}^{(2)}(\mathrm{S})$, in which we are interested in,
receives no SUSY contributions, because the gauginos couple via a
gauge coupling and, thus, contribute merely to $g^4$ terms of
$\hat{\gamma}^{(2)}(\mathrm{S})$. The generic result of
Eq.~\eqref{eqn_general_two-loop_results} is altered in a
SUSY theory only by a factor of $1/2$ due to the different
normalization of the Yukawa couplings.  We obtain
\begin{subequations}
  \begin{align}
     \hat{\gamma}_{aa}^{(2)}(\mathrm{S})  \Big|_{\text{SUSY}} &= - \frac{1}{(
4\pi)^4} g^2 \xi \xi' \GenGauge{a}{c}{A} \YukaInvGauge{c}{d}
\GenGauge{d}{a}{A}  + \mathcal{\tilde{R}}_{a a}, \\
  \beta^{(2)}(v_{a}) \Big|_{\text{SUSY}} &= \gamma_{aa}^{(2)}(\mathrm{S}) v_{a}
- \frac{1}{( 4 \pi)^4} g^2 \xi \xi' \GenGauge{a}{c}{A} \YukaInvGauge{c}{d}
\GenGauge{d}{a}{A}  v_{a} + \mathcal{\tilde{R}}_{a a} v_a.
  \end{align}
\end{subequations}
Note that $\mathcal{R}$ is altered to $\mathcal{\tilde{R}}$ because the term
$\sim g^2 Y Y^{\dagger}$ leads to a $g^4$ contribution due to the Gaugino
coupling.

\section{Application to Concrete SUSY Models}
\label{sec_4_results}
The aim of this section is to apply the results from Section~\ref{sec_3_methods}
to the MSSM and non-minimal supersymmetric models. We discuss the validity of
Eq.~\eqref{eqn_VEV_diff} and provide new, explicit results for the NMSSM and
\ESSM. For this application, we need to generalize our results to product gauge
groups, see Ref.~\cite{Machacek:1983tz}, and we use model-specific expressions for
the Yukawa couplings.

\subsection{MSSM}
The MSSM~\cite{Haber:1984rc} contains two Higgs doublets $H_u, H_d$ with
opposite hypercharge $Y_{H_u}/2=-Y_{H_d}/2=1/2$. Our calculations are based upon
the following superpotential\footnote{The dot product $A \cdot B= \epsilon_{ij}
A_i B_j$ denotes the $SU(2)$ invariant product with antisymmetric $\epsilon_{ij}$.}
  \begin{align}
    \mathcal{W}_{\text{MSSM}} = \mu H_d \cdot H_u - y_{ij}^{e}H_d
    \cdot L_i \bar{E}_j - y_{ij}^{d}H_d\cdot Q_i \bar{D}_j -
    y_{ij}^{u}Q_i\cdot H_u \bar{U}_j.
    \label{MSSM_superpotential}
  \end{align}
Applying Eq.~\eqref{eqn_vev_beta_one-loop_susy} is in agreement with
the known results for the divergent renormalization constants (and
equivalent the $\beta$-functions)
\begin{subequations}
 \begin{align}
  \frac{\beta_{\text{MSSM}}^{(1)}(v_{u})}{v_u} &= \frac{1}{(4 \pi)^2}
\left[ \left(1-\xi  + 2 \xi \xi'\right) \left( \frac{3}{20} g_1^2 +
\frac{3}{4} g_2^2 \right) - N_c \tr \left( y^{u} y^{u \dagger} \right)
\right] \nonumber \\
    &=\gamma^{(1)}_{uu} + \frac{1}{(4 \pi)^2} \xi \xi' \left( \frac{3}{10}
g_1^2 + \frac{3}{2} g_2^2 \right) , \\
  \frac{\beta_{\text{MSSM}}^{(1)}(v_{d})}{v_d} &= \frac{1}{(4 \pi)^2}
\left[ \left(1-\xi  + 2 \xi \xi'\right) \left( \frac{3}{20} g_1^2 +
\frac{3}{4} g_2^2 \right) - N_c\tr \left( y^{d} y^{d \dagger} \right)
- \tr \left( y^{e} y^{e \dagger} \right) \right] \nonumber \\
    &= \gamma^{(1)}_{dd} + \frac{1}{(4 \pi)^2} \xi \xi' \left(
\frac{3}{10} g_1^2 + \frac{3}{2} g_2^2 \right) . 
 \end{align}
  \label{eqn_vev_CT_1-loop_MSSM}
\end{subequations}
Here $g_{Y}$ and $g_{2}$ are the $U(1)_Y$ and $SU(2)_L$ gauge
couplings, and $g_1$ is GUT-normalized $g_1=\sqrt{5/3}g_Y$. $N_c$ is
the number of colours.

The quantity $\tan \beta$ is defined as
\begin{align}
  \tan\beta = \frac{v_{u}}{v_{d}} ,
\end{align}
and renormalization yields
\begin{subequations}
  \begin{align}
 \delta \tan\beta^{(1)} &= \tan\beta \left( \frac{\delta v_{u}^{(1)}}{v_{u}} -
\frac{\delta v_{d}^{(1)}}{v_{d}} \right) ,\\
 \beta^{(1)}(\tan\beta) &= \tan\beta \left( \gamma^{(1)}_{uu}
-\gamma^{(1)}_{dd} +  \hat{\gamma}^{(1)}_{uu} -\hat{\gamma}^{(1)}_{dd} \right) .
  \end{align}
\end{subequations}
There are two cancellations in this difference. First, the
contribution from the $\hat{\gamma}$'s, equivalent to $\delta \bar{v}$
in Eq.~\eqref{DeltaVDecomposition}, cancels. The reason is that both
doublets have the same $SU(2)_L$ and $U(1)_Y$ quantum numbers, up to a
sign. Hence, Eq.~\eqref{eqn_VEV_diff} is valid at the one-loop level
and for the $\beta$-function the simplified result
\begin{align}
 \beta_{\text{MSSM}}^{(1)}(\tan\beta) = \tan\beta \left( \gamma^{(1)}_{uu}
-\gamma^{(1)}_{dd}  \right)
\end{align}
holds. Second, even within this difference, the gauge coupling terms
cancel, leaving
\begin{align}
  \frac{\beta_{\text{MSSM}}^{(1)}(\tan\beta)}{\tan\beta} = - \frac{1}{(4
\pi)^2}
   \left[ N_c \tr \left( y^{u} y^{u \dagger} \right)  - N_c \tr \left( y^{d}
y^{d \dagger} \right) -  \tr \left( y^{e} y^{e \dagger} \right)  \right].
\end{align}
Obviously, both cancellations are group theoretical coincidences. As a remark,
these results also provide further insight into the accidental gauge
independence of $\tan \beta$  as discussed in Ref.~\cite{Freitas:2002pe}. Going
away from $R_\xi$-gauges, $\tan \beta$ becomes gauge dependent at the
one-loop level
\cite{Freitas:2002pe,Baro:2008bg}.

At two-loop level, the $\beta$-functions for the up and down type VEVs are given
by
\begin{subequations}
 \begin{align}
   \frac{\beta_{\text{MSSM}}^{(2)}(v_{u})}{v_u} &= \gamma_{u u}^{(2)} -
\frac{1}{(4 \pi)^4} \xi \xi' \left( \frac{3}{10} g_1^2 + \frac{3}{2} g_2^2
\right) \left[ N_c \tr \left( y^{u} y^{u \dagger} \right) \right]  +
\mathcal{R}_u , \label{eqn_beta_vev_u} \\
   \frac{\beta_{\text{MSSM}}^{(2)}(v_{d})}{v_d} &= \gamma_{d d}^{(2)} -
\frac{1}{(4 \pi)^4} \xi \xi' \left( \frac{3}{10} g_1^2 + \frac{3}{2} g_2^2
\right) \left[N_c \tr \left( y^{d} y^{d \dagger} \right) +\tr \left( y^{e} y^{e
\dagger} \right) \right]  + \mathcal{R}_d  \label{eqn_beta_vev_d} .
 \end{align}
\end{subequations}
Here $\mathcal{R}_{u,d}$ represent all contributions from $\delta
\hat{Z}^{(2)} \sim g^4$ which we have not considered. However, it is
clear that in the MSSM
$\mathcal{R}_{u} = \mathcal{R}_{d}$ holds because the Higgs doublets have (up
to a sign) the same quantum numbers with respect to all gauge groups. Our
results are in agreement with Ref.~\cite{Yamada:2002nu} if one simplifies the
generation matrices $y_{ij}^{d/u/e}$ to complex numbers $y^{d/u/e}$.

The above result now implies that the $g_{2}^{2}$ and $g_{1}^{2}$ proportional
parts of the divergent $\delta\tan\beta$ are not zero if the Yukawa
couplings are different. Instead, we can write 
\begin{align}
 \frac{\beta_{\text{MSSM}}^{(2)}(\tan\beta)}{\tan\beta} &= \gamma^{(2)}_{uu}
 -\gamma^{(2)}_{dd} + \frac{1}{(4 \pi)^2} \xi \xi' \left( \frac{3}{10}
g_1^2 + \frac{3}{2} g_2^2 \right)
\frac{\beta_{\text{MSSM}}^{(1)}(\tan\beta)}{\tan\beta} 
\label{eqn_beta_tan_two-loop_MSSM}
\end{align}
The second term is equivalent to a violation of Eq.~\eqref{eqn_VEV_diff} at
$\mathcal{O}(g^2 Y Y^\dagger)$. Note that $\mathcal{R}_{u} -
\mathcal{R}_{d}$ vanishes in the MSSM as remarked earlier. 


\subsection{NMSSM}
The NMSSM is a non-minimal SUSY model that attempts to solve the $\mu$
problem of the MSSM by introducing an additional gauge singlet $S$,
which has a non-zero VEV $v_s$. The modified superpotential, see for
example Ref.~\cite{PhysRevD.39.844}, reads as
\begin{align}
 \mathcal{W}_{\text{NMSSM}} = \mathcal{W}_{\text{MSSM}}(\mu=0) + \lambda S H_d
\cdot  H_u + \frac{1}{3} \kappa S S S + \frac{1}{2} \mu_s S S + \zeta S .
\end{align}
The one-loop $\beta$-functions are given by
\begin{subequations}
  \begin{align}
    \frac{\beta_{\text{NMSSM}}^{(1)}(v_s)}{v_s} &= - \frac{1}{(4 \pi)^2} 2
\left(|\lambda|^2 + |\kappa|^2 \right) ,\\
  \frac{ \beta_{\text{NMSSM}}^{(1)}(v_{u,d})}{v_{u,d}} &=
\frac{\beta_{\text{MSSM}}^{(1)}(v_{u,d})}{v_{u,d}}- \frac{1}{(4 \pi)^2}
|\lambda|^2 .
  \end{align}
\end{subequations}
The consequences of the singlet $S$ for the Higgs VEVs are a change in the
anomalous dimensions $\gamma_{uu}$, $\gamma_{dd}$ due to the additional Yukawa
coupling $\lambda$, whereas the $\hat{\gamma}_{uu}$, $\hat{\gamma}_{dd}$ are
unchanged as $S$ is a gauge singlet. In addition, the quantities
$(\sqrt{\hat{Z}_s}-1)$ and $\hat{\gamma}_{ss}$ vanish, because an additional
counterterm for $v_s$ is
forbidden by the rigid invariance for $S$. Therefore, the one-loop $\beta$-function for $\tan \beta$ reads 
\begin{align}
 \beta_{\text{NMSSM}}^{(1)}(\tan \beta) = \beta_{\text{MSSM}}^{(1)}(\tan
\beta),
\end{align}
and Eq.~\eqref{eqn_VEV_diff} holds.

Similarly, the changes in $\beta^{(2)}(v_{u,d,s})$ are
\begin{subequations}
 \begin{align}
  \frac{ \beta_{\text{NMSSM}}^{(2)}(v_{u})}{v_u} &= \gamma_{u u}^{(2)} -
\frac{1}{(4 \pi)^4} \xi \xi' \left( \frac{3}{10} g_1^2 + \frac{3}{2} g_2^2
\right) \left[ N_c \tr \left( y^{u} y^{u \dagger} \right) + |\lambda|^2 \right]
+ \mathcal{R}_u , \label{eqn_beta_vev_u_nmssm} \\
  \frac{ \beta_{\text{NMSSM}}^{(2)}(v_{d})}{v_d} &= \gamma_{d d}^{(2)} -
\frac{1}{(4 \pi)^4} \xi \xi' \left( \frac{3}{10} g_1^2 + \frac{3}{2} g_2^2
\right) \left[ N_c \tr \left( y^{d} y^{d \dagger} \right) +\tr \left( y^{e}
y^{e \dagger} \right) + |\lambda|^2 \right] + \mathcal{R}_d ,
  \label{eqn_beta_vev_d_nmssm} \\
  \frac{ \beta_{\text{NMSSM}}^{(2)}(v_{s})}{v_s} &= \gamma_{s s}^{(2)} .
  \label{eqn_beta_vev_s_nmssm}
 \end{align}
\end{subequations}
Again, $\mathcal{R}_u = \mathcal{R}_d$ because the gauge groups of NMSSM and
MSSM are identical.

For the $\tan \beta$ two-loop $\beta$-function we obtain a result reminiscent
of the MSSM Eq.~\eqref{eqn_beta_tan_two-loop_MSSM}
\begin{align}
 \frac{ \beta_{\text{NMSSM}}^{(2)}(\tan\beta)}{\tan\beta} &=  \gamma^{(2)}_{uu}
-\gamma^{(2)}_{dd} + \frac{1}{(4 \pi)^2} \xi \xi' \left( \frac{3}{10}
g_1^2 + \frac{3}{2} g_2^2 \right) \frac{
\beta_{\text{MSSM}}^{(1)}(\tan\beta)}{\tan \beta}  .
\label{eqn_beta_tan_two-loop_NMSSM}
\end{align}
Note that $\gamma$ and $\mathcal{R}$ refer to the NMSSM quantities,
i.e.\ they differ from the MSSM quantities. Nevertheless, the difference
$\mathcal{R}_{u} - \mathcal{R}_{d}$ still vanishes in the NMSSM.


\subsection{\ESSM}  
The \ESSM\ is based on the direct product gauge group $SU(3)_c \otimes
SU(2)_L \otimes U(1)_Y \otimes U(1)_N$. Among the several Higgs doublet fields
$\{H_{u,i} \}_{i=1,2,3}$ and $\{ H_{d,i} \}_{i=1,2,3}$  only the third
generation (i.e.\ $i=3$) of up- and down-type Higgs acquire a non-zero VEV
$v_{u,d}$ \cite{King:2005jy}. The $U(1)_N$ charges of the Higgs fields are given
as $N_{H_{u,i}}/2 = -2$ and $N_{H_{d,i}}/2 =-3 $. Similarly, there
are three generations of SM-group singlets $S_i$, which carry $U(1)_N$
charge $N_{S_{i}}/2=5$, and only $S_3$ has a non-zero VEV
$v_s$. Following Ref.~\cite{King:2005jy} and using the same convention
as in the MSSM, the approximated \ESSM\ superpotential can be written
as
\begin{align}
 \mathcal{W}_{\text{\ESSM}} \approx  &- y_{ij}^{e} H_{d,3} \cdot L_i \bar{E}_j
- y_{ij}^{d} H_{d,3} \cdot Q_i \bar{D}_j -  y_{ij}^{u} Q_i\cdot H_{u,3}
\bar{U}_j \label{ESSM_superpotential} \\
  &+ \lambda_i S_3 H_{d,i} \cdot H_{u,i} + \kappa_i S_3 X_i \bar{X}_i.\notag
\end{align}
The superfields $X_i$, $\bar{X}_i$ describe exotic colored matter and
transform as triplet/anti-triplet under $SU(3)$, as singlet under
$SU(2)_L$ and have $U(1)_Y$ and $U(1)_N$ quantum numbers $Y_{X_i}/2=-
1/3$, $Y_{\bar{X}_i}/2= 1/3$, $N_{X_i}/2=-2$, $N_{\bar{X}_i}/2=-3$.

For $\beta(v_{u,d})$ and $\beta(v_s)$ we obtain a similar expression as in
the MSSM/NMSSM with one profound difference: The $U(1)_N$ charge difference
leads to a non-vanishing contribution. The Casimir eigenvalue leads to
$ (N/2)^2$ contributions and, thus, the one-loop $\beta$-functions read
\begin{subequations}
  \begin{align}
    \frac{\beta_{\text{\ESSM}}^{(1)}(v_s)}{v_s} &= \frac{1}{(4 \pi)^2} \left[
g_N^2 \left(1-\xi + 2\xi \xi' \right) \left(\frac{N_S}{2} \right)^2 -
2 \tr \left(\lambda \lambda^\dagger \right) - N_c \tr \left( \kappa
\kappa^\dagger \right) \right] ,\\
   \frac{ \beta_{\text{\ESSM}}^{(1)}(v_{u,d})}{v_{u,d}} &=
\frac{\beta_{\text{MSSM}}^{(1)}(v_{u,d})}{v_{u,d}}+ \frac{1}{(4 \pi)^2} \left[
g_N^2 \left(1-\xi + 2\xi \xi' \right)  \left(\frac{N_{H_u,H_d}}{2} \right)^2 -
|\lambda_3|^2 \right].
  \end{align}
\end{subequations}
In contrast to the NMSSM, the SM-singlet $S_3$ has a non-vanishing $\hat{\gamma}_{ss}$-contribution due to the $U(1)_N$ gauge coupling.

For $\tan \beta$ those one-loop results yield
  \begin{align}
 \frac{\beta_{\text{\ESSM}}^{(1)}(\tan\beta) }{\tan\beta} &= \frac{1}{(4
\pi)^2} \Bigg\{ g_{N}^{2} \left(1-\xi  + 2\xi \xi'\right) \left[ \left(
\frac{N_{H_u}}{2} \right)^2 - \left( \frac{N_{H_d}}{2} \right)^2 \right]
\label{eqn_tan_beta_one-loop_ESSM}\\
    &\phantom{= \frac{1}{(4\pi)^2} \Bigg\{}
 - \left[ N_c \tr \left( y^{u} y^{u \dagger} \right)  -
N_c \tr \left( y^{d} y^{d \dagger} \right) -  \tr \left( y^{e} y^{e \dagger}
\right)  \right] \Bigg\}  .\nonumber 
  \end{align}
As a consequence, both MSSM one-loop cancellations do not occur in the
\ESSM: neither the $\hat{\gamma}$-terms nor the gauge-coupling terms
within $\gamma$ drop out
because of the different Higgs $U(1)_N$ charges.
Eq.~\eqref{eqn_tan_beta_one-loop_ESSM} is in agreement with the result of
Ref.~\cite{Athron:2012pw} obtained from finiteness of the renormalized
Yukawa couplings. 

As a further application, we present the two-loop results for
$\beta(\tan\beta)$ in the \ESSM. First, the VEV $\beta$-functions read at
two-loop level
\begin{subequations}
  \begin{align}
 \frac{\beta_{\text{\ESSM}}^{(2)}(v_{s})}{v_{s}} &= \gamma_{s s}^{(2)} -
\frac{1}{(4 \pi)^4} 2 \xi \xi' \left(\frac{N_S}{2} \right)^2 g_N^2 \left[ 2 \tr
\left(\lambda \lambda^\dagger \right) + N_c \tr \left( \kappa \kappa^\dagger
\right) \right]  + \mathcal{R}_s , \\
  \frac{\beta_{\text{\ESSM}}^{(2)}(v_{u})}{v_{u}} &= \gamma_{u u}^{(2)} -
\frac{1}{(4 \pi)^4} \xi \xi' \left( \frac{3}{10} g_1^2 + \frac{3}{2} g_2^2  + 2
\left(\frac{N_{H_u}}{2}\right)^2 g_N^2\right) \left[ N_c \tr \left( y^{u} y^{u
\dagger} \right) + |\lambda_3|^2 \right]  + \mathcal{R}_u 
, \label{eqn_beta_vev_u_e6ssm} \\
  \frac{ \beta_{\text{\ESSM}}^{(2)}(v_{d})}{v_{d}} &= \gamma_{d d}^{(2)} -
  \frac{1}{(4 \pi)^4} \xi \xi' \left( \frac{3}{10} g_1^2 + \frac{3}{2} g_2^2 + 2
    \left(\frac{N_{H_d}}{2}\right)^2 g_N^2 \right) \label{eqn_beta_vev_d_e6ssm}\\
  &\hphantom{=\;\gamma_{d d}^{(2)} - \frac{1}{(4 \pi)^4} \xi \xi'}
  \times\left[ N_c \tr \left( y^{d} y^{d
        \dagger} \right) +\tr \left( y^{e} y^{e \dagger}  \right) + |\lambda_3|^2
  \right] + \mathcal{R}_d.\notag
  \end{align}
\end{subequations}
Unlike the MSSM and NMSSM case, $\mathcal{R}_u \neq \mathcal{R}_d$ in
the \ESSM\ as the Higgs doublets have different $U(1)_N$ quantum
numbers.  Furthermore, the contributions from kinetic mixing of the
$U(1)_Y$ and $U(1)_N$ groups \cite{Fonseca:2011vn} are not relevant
for the $\mathcal{O}(g^2 Y Y^\dagger)$ contributions we have
explicitely given.

Second, for $\tan \beta$ follows at two loop
\begin{subequations}
  \begin{align}
 \frac{ \beta_{\text{\ESSM}}^{(2)}(\tan\beta)}{\tan\beta} &= \gamma^{(2)}_{uu}
-\gamma^{(2)}_{dd} + \frac{1}{(4
\pi)^2} \xi \xi' \left( \frac{3}{10} g_1^2 + \frac{3}{2} g_2^2  \right)
 \frac{\beta^{(1)}_{\text{MSSM}} (\tan \beta)}{\tan\beta} 
\label{eqn_beta_tan_two-loop_ESSM} \\
  &-\frac{2}{(4\pi)^4} \xi \xi' g_N^2 \Bigg\{ \left( \frac{N_{H_u}}{2}\right)^2 
 N_c \tr \left( y^{u} y^{u \dagger} \right) - \left(\frac{N_{H_d}}{2}\right)^2 
  \left[ N_c \tr \left( y^{d} y^{d \dagger} \right) +\tr \left( y^{e}
  y^{e\dagger} \right) \right] \Bigg\}  \nonumber \\
  &-\frac{2}{(4\pi)^4} \xi \xi' g_N^2 \left[ \left( \frac{N_{H_u}}{2}\right)^2 -
\left( \frac{N_{H_d}}{2}\right)^2 \right] |\lambda_3|^2  + \mathcal{R}_u -
\mathcal{R}_d . \nonumber
  \end{align}
\end{subequations}
Note the structure of Eq.~\eqref{eqn_beta_tan_two-loop_ESSM}: the second term
corresponds to the MSSM $\mathcal{O}(g^2 YY^\dagger)$ term from
Eq.~\eqref{eqn_beta_tan_two-loop_MSSM}, whereas the second and third line
represent further violations of Eq.~\eqref{eqn_VEV_diff} due to the $U(1)_N$
couplings. 

\section{Conclusions}
We computed the \MSbar / \DRed\ $\beta$-function for VEVs in general gauge
theories and general SUSY gauge theories (Wess-Zumino gauge) up to
Yukawa-enhanced two-loop contributions. These results complement the
$\beta$-functions of
Refs.~\cite{Machacek:1983tz,Machacek:1983fi,Machacek:1984zw,Martin:1993zk,Yamada:1994id,Jack:1994kd}. In
addition, we provided the $\beta$-functions for  $\tan \beta$ in
the MSSM, NMSSM, and \ESSM\ up to this order in
general $R_\xi$ gauge. These $\beta$-functions
are required in renormalization group studies of spontaneously broken gauge
theories, and they can be implemented in computer codes, e.g.\ in
spectrum-generator generators like \emph{SARAH}
\cite{Staub:2010jh,Staub:2011dp} 
or many existing MSSM or NMSSM spectrum generators. 

Our results have been obtained by using the elegant approach of
Ref.~\cite{Kraus:1995jk}, which is interesting in its own right. In the past,
this approach has been applied in more abstract contexts, but we have shown that
it also facilitates calculations and provides qualitative understanding. We
therefore close by summarizing this approach and its consequences.
\begin{itemize}
 \item The VEVs $v$ are promoted to background fields. As a consequence,
$R_\xi$ gauge fixing can be formulated without breaking global gauge
invariance.
\item The renormalization of the VEVs is completely determined by the
  field renormalization of the fields and background fields. The VEV
  counterterm can be expressed in terms of the dimension-zero field
  renormalization constants $\sqrt{Z}$ and $\sqrt{\hat{Z}}$, and
  $(\sqrt{\hat{Z}}-1) \VevGauge{}$ replaces $\delta \bar{v}$ from
  Eq.~\eqref{DeltaVDecomposition}.
  \item The $\beta$-function of the VEV is similarly composed of the
    anomalous dimensions 
$\gamma(\mathrm{S})$ and $\hat{\gamma}(\mathrm{S})$ of the fields and
background fields, see Eq.\ \eqref{eqn_vev_beta_general}.
\item The approach leads to additional information since the new
  renormalization constant $\sqrt{\hat{Z}}$ appears in the Lagrangian
  in a well-defined manner. As a consequence, its computation requires
  to evaluate the Green function
  $\Gamma_{\BrstBackGauge{a},K_{\ScaGauge{b}}}$, which is unphysical
  but very simple to compute.
  \item The vertices in Fig.\ \ref{fig_BRS-source_ghost_scalar_vertex},
    \ref{fig_BRS_background-field_anti-ghost_scalar_vertex}
    contributing to $\Gamma_{\BrstBackGauge{a},K_{\ScaGauge{b}}}$ loop
    corrections are dictated by BRS-invariance and, thus, are
    restricted to gauge couplings. In particular, non-trivial
    $\BrstBackGauge{}$-vertices can only arise from
    Eq.~\eqref{eqn_def_fix-gh} and are thus linked to the gauge fixing
    term. Gauges such as Landau gauge, where $F^A$ is independent of
    $\ScaBackGauge{}$, do not lead to such $\BrstBackGauge{}$-vertices
    and have $\sqrt{\hat{Z}}=1$ and $\delta \bar{v}=0$.
  \item The cancellation in $\tan \beta$,
Eq.~\eqref{eqn_VEV_diff}, as observed in
Refs.~\cite{Dabelstein:1994hb,Chankowski:1992er}, is a group theoretic
coincidence and can be understood from the
general expression for $\hat{\gamma}$. 
\end{itemize}

\section*{Acknowledgments} D.S.\ thanks M.\ Beneke, A.\ Dabelstein,
W.\ Hollik, D.R.T.\ Jones and H.\ Rzehak for discussions and comments
related to this work.  A.V.\ is supported by the German Research
Foundation DFG, RTG 1504.

\bibliographystyle{h-physrev}     
\footnotesize{\bibliography{Literatur}}      

\end{document}